\documentclass{elsart}

\usepackage{latexsym}
\usepackage{psfig}

\newcommand{\beq}{\begin{equation}}
\newcommand{\eeq}{\end{equation}}
\newcommand{\ba}{\begin{array}}
\newcommand{\ea}{\end{array}}
\newcommand{\bea}{\begin{eqnarray}}
\newcommand{\eea}{\end{eqnarray}}
\newcommand{\bc}{\begin{center}}
\newcommand{\ec}{\end{center}}
\newcommand{\bt}{\begin{table}}
\newcommand{\et}{\end{table}}

\newcommand{\la}[1]{\label{#1}}
\newcommand{\p}{\partial}

\newcommand{\pp}[2]{{\partial #1 \over \partial #2}}
\newcommand{\ppn}[3]{{\partial^{#1} #2 \over \partial #3^{#1}}}

\newcommand{\rf}[1]{(\ref{#1})}
\newcommand{\beqno}{\begin{displaymath}}
\newcommand{\eeqno}{\end{displaymath}}

\newcommand{\been}{\begin{enumerate}}
\newcommand{\een}{\end{enumerate}}

\newcommand{\sn}{{\rm sn}}
\newcommand{\cn}{{\rm cn}}
\newcommand{\dn}{{\rm dn}}

\newcounter{saveeqn}

\newcommand{\alpheqn}{\setcounter{saveeqn}{\value{equation}}
\stepcounter{saveeqn}\setcounter{equation}{0}
\renewcommand{\theequation}{\mbox{\arabic{saveeqn}\alph{equation}}}}

\newcommand{\resetalpheqn}{\setcounter{equation}{\value{saveeqn}}
\renewcommand{\theequation}{\arabic{equation}}}

\newcommand{\primeeqn}[1]{\setcounter{saveeqn}{\value{equation}}
\renewcommand{\theequation}{\ref{#1}$'$}}

\newcommand{\resetprimeeqn}{\setcounter{equation}{\value{saveeqn}}
\renewcommand{\theequation}{\arabic{equation}}}

\journal{Physics Letters A}
\begin{document}

\begin{frontmatter}

\title{Stability of exact solutions of the defocusing nonlinear 
   Schr\"odinger equation with periodic potential in two dimensions}

\author{Bernard Deconinck\thanksref{1}},
\author{Bela A. Frigyik},
\author{J. Nathan Kutz\thanksref{2}}

\thanks[1]{Corresponding author 
(deconinc@amath.washington.edu, phone: (206) 685-2971, fax: (206) 685-1440).  
Acknowledges support 
from the National Science Foundation (DMS-0071568).}
\thanks[2]{Acknowledges support from the National Science Foundation (DMS-0092682).}

\address{Department of Applied Mathematics, 
  University of Washington, Seattle, WA 98195-2420\\
 {\bf A higher resolution version of this preprint is available at:
  http://www.amath.washington.edu/~kutz/research.html}}

\begin{abstract}
The cubic nonlinear Schr\"odinger equation with repulsive nonlinearity and
elliptic function potential in two-dimensions models a repulsive
dilute gas Bose--Einstein condensate in a lattice potential.  
A family of exact stationary solutions is presented and its stability is
examined using analytical and numerical methods.
All stable trivial-phase solutions are off-set from the zero level.
Our results imply that a large number of
condensed atoms is sufficient to form a stable, periodic condensate. 
\end{abstract}


\end{frontmatter}


\section{Introduction}

The cubic nonlinear Schr\"odinger equation with attractive or repulsive
nonlinearity and a potential is used as a mean-field model for the dynamics of
a dilute-gas Bose Einstein condensate (BEC) \cite{gross,pitaevskii}. In this
case, the equation is often referred to as the Gross-Pitaevskii equation.
These BECs are of interest in both the theoretical and experimental physics
community: they are examples of macroscopic quantum phenomena which display
phase coherence~\cite{anderson1,hagley1,hagley2,chio} and which have potential
applications in such diverse areas as quantum logic \cite{brennen},
matter-wave diffraction \cite{diffraction}, and matter-wave transport
\cite{transport}.

After rescaling of the physical variables, the governing equation is 
\beq
\label{eqn:nls}
 i \pp{}{t}\psi(\vec{x},t)= -\frac{1}{2}\Delta\psi(\vec{x},t) + 
 \alpha \left|\psi(\vec{x},t)\right|^2 \psi(\vec{x},t) 
        + V(\vec{x}) \psi(\vec{x},t),
\eeq
where $\Delta$ denotes the Laplacian operator, $\psi(\vec{x},t)$ is the
macroscopic wave function of the condensate in one, two or three dimensions,
with $\vec{x}$ defined as $x$, $(x,y)$ or $(x,y,z)$ respectively.  The
real-valued function $V(\vec{x})$ is an experimentally generated macroscopic
potential. The parameter $\alpha$ determines whether (\ref{eqn:nls}) is
repulsive ($\alpha=1$, defocusing nonlinearity), or attractive, ($\alpha=-1$, 
focusing nonlinearity).
Note that for BEC applications, both signs of $\alpha$ are relevant.

Many BEC experiments use harmonic confinement, but recently there has been
interest in confinement of repulsive BECs using standing light waves, resulting
in a sinusoidal potential in the Nonlinear Schr\"odinger equation. So far, most
interest has focussed on the quasi-one-dimensional regime, in which the
longitudinal dimension of the condensate is far greater than the transverse
dimensions, which are themselves of the order of the healing length of the
condensate. In this case, the governing mean-field equation is \rf{eqn:nls},
with $\alpha=1$, in one dimension, and the experimentally generated potential
$V(x)=-V_0~\sin^2(x)$. In \cite{becprl1,becpre1}, a number of exact solutions of
this equation was constructed and their stability was investigated. In 
fact, a potential more general than a sinusoidal potential was considered: 
\beq\la{eqn:pot1}
V(x)=-V_0~\sn^2(x,k),
\eeq
where $\sn(x,k)$ denotes the Jacobian elliptic sine function, with elliptic
modulus $k$. In the limit $k=0$, this potential reduces to a sinusoidal potential.
It is important to realize that for values of $k$ up to $k=0.9$, the shape of
the potential is virtually indistinguishable from a sinusoidal one. On the other
hand, considering values of $k$ close to $1$, $i.e.,$ $k>0.999$ gives periodic
potentials with well-separated peaks, allowing the study of BEC dynamics in an
entirely new regime. 

\begin{figure}[tb]
\centerline{\psfig{figure=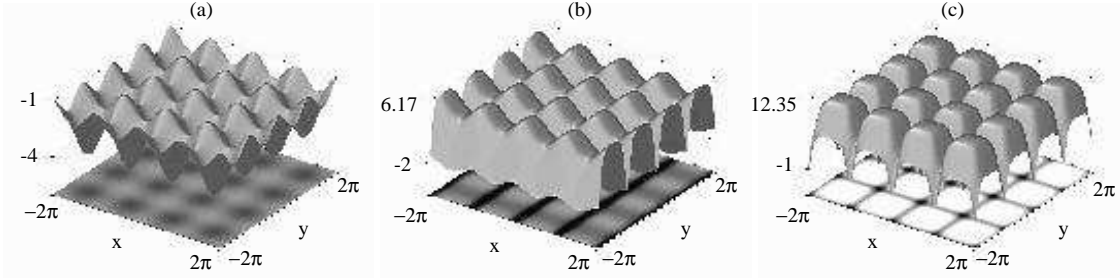,width=150mm,silent=}}
\begin{center}
\caption{ \label{fig:pots} Various lattice potentials. For all figures,
$A_1=A_2=B_1=B_2=1$. For (a), $k_1=k_2=0$, $m_1=m_2=1$; for (b), $k_1=0.999$,
$k_2=0$, $m_1=2K(0.999)/\pi$, $m_2=1$. Finally, for (c) $k_1=k_2=0.999$ and
$m_1=m_2=2K(0.999)/\pi$.}
\end{center}
\end{figure}

Currently, no experiments are being performed where a BEC is trapped in a
higher-than-one-dimensional periodic potential. However, the interest in the
applications mentioned above strongly suggests that these experiments may
eventually take place.  Already, there are theoretical investigations of BECs
in multidimensional lattice potentials \cite{jaksch}, suggesting the
realization of such experiments.  Although motivated by the developments in
BECs, in this paper we consider (\ref{eqn:nls}) with repulsive nonlinearity in
two dimensions in its own right.  Thus we consider
\primeeqn{eqn:nls}
\beq\la{eqn:nls2d}
i\pp{\psi}{t}=-\frac{1}{2}\left(\ppn{2}{\psi}{x}+\ppn{2}{\psi}{y}\right)
+\left|\psi\right|^2 \psi+V(x,y) \psi, 
\eeq
\resetprimeeqn
with $\psi=\psi(x,y,t)$. 
As in one dimension~\cite{becprl1,becpre1}, 
making the proper choices for the potential allows the
construction of a large class of exact solutions. 
The family of potentials considered is
\begin{eqnarray}
\nonumber
V(x,y)&=&-\left(A_1 \sn^2(m_1 x,k_1)+B_1\right)
\left(\!A_2 \sn^2(m_2 y,k_2)+B_2\right)+\\\label{eqn:pot2}
&&+m_1^2 k_1^2 \sn^2(m_1 x,k_1)+m_2^2 k_2^2 \sn^2(m_2
y,k_2). 
\end{eqnarray}
Here $A_1, A_2, B_1, B_2, m_1$ and $m_2$ are real constants. The two elliptic
moduli $k_1$ and $k_2$ are in the interval $[0,1]$. 
The first term is a straightforward generalization of the one-dimensional
potential (\ref{eqn:pot1}), with an additive constant.  The other terms
are incorporated to facilitate the construction of exact solutions.  Although
this exact expression for the potential is necessary to allow the construction
of exact solutions, it is the qualitative features, $i.e.,$ its periodicity and
amplitude, that are most important. Numerical and analytical results throughout
this paper demonstrate that the behavior of a solution in a lattice 
potential is largely independent of the quantitative features of the potential.
Figure \ref{fig:pots} displays the potential \rf{eqn:pot2} for various
values of its parameters. 
%
%
%
%
In the next section, a family of exact solutions to \rf{eqn:nls2d} is given and
their linear stability is discussed. Numerical results for various
representative solutions are given in Section \ref{sec:num}. The last section
concludes the paper with a brief summary of the relevant mathematical
and physical results. 

\section{Analytical Results} 

A judicious choice of the potential allows for the cancellation of the
nonlinear term in \rf{eqn:nls2d} so that exact solutions can be constructed.
Of course, one can always find a suitable potential by solving \rf{eqn:nls2d}
for $V(x,y)$, given a certain $\psi(x,y,t)$. However, this generically results
in time-dependent potentials and is hence not of interest. The kind of
potentials we seek are two-dimensional generalizations of the one-dimensional
potential \rf{eqn:pot1}. This one-dimensional potential is periodic. Its
two-dimensional generalizations should be periodic in both its spatial
variables. Furthermore, it is desirable that for a given potential more than
one exact solution exists.

A derivation of the exact solutions is not presented in this paper. Rather we
state the solution formulas and discuss them.  For the potential \rf{eqn:pot2},
a class of exact solutions is given by   
\beq\la{eqn:ans} 
\psi(x,y,t)=r_1(x) r_2(y) e^{i\theta_1(x)+i\theta_2(y)-i\omega t}, 
\eeq 
with 
\alpheqn 
\bea\la{eq:amphase} 
r_1^2(x)=A_1~\sn^2(m_1 x,k_1)+B_1,~~
\theta_1(x)=c_1~\int_0^x\frac{dz}{A_1~\sn^2(m_1 z,k_1)+B_1},\\
r_2^2(y)=A_2~\sn^2(m_2 y,k_2)+B_2,~~
\theta_2(y)=c_2~\int_0^y\frac{dz}{A_2~\sn^2(m_2 z,k_2)+B_2},  
\eea
\resetalpheqn 
and  
\alpheqn 
\bea 
\omega&=&\frac{1}{2} m_1^2\left(1+k_1^2+k_1^2 \frac{B_1}{A_1}\right)+ 
\frac{1}{2} m_2^2\left(1+k_2^2+k_2^2 \frac{B_2}{A_2}\right),\\
c_1^2&=&m_1^2\frac{B_1}{A_1}\left(A_1+B_1\right)\left(A_1+k_1^2 B_1\right),\\
c_2^2&=&m_2^2\frac{B_2}{A_2}\left(A_2+B_2\right)\left(A_2+k_2^2 B_2\right).
\eea 
\resetalpheqn 
Choosing the 
parameters $A_1 A_2$, $B_1/A_1$, $B_2/A_2$,
$m_1$, $m_2$, $k_1$ and $k_2$ determines the potential. 
Thus the solution class given by
(\ref{eq:amphase}-b) is a one-parameter family of solutions, with free
parameter $A_1/A_2$.   Existence of these solutions requires $r_1^2\geq 0,
c_1^2\geq 0$ and  $r_2^2 \geq 0, c_2^2\geq 0$. This imposes conditions on the
parameters: $A_i\geq 0, B_i \geq 0$ or $B_i\geq 0, -A_i\leq B_i
\leq-A_i/k_i^2$, where $i=1,2$.

From (\ref{eq:amphase}-b), it follows that $r_1(x)$ is periodic in $x$ with
period $2K(k_1)/m_1$, where $K(k)=\int_0^{\pi/2} d z/\sqrt{1-k_1^2 \sin^2(z)}$,
and $r_2(y)$ is periodic in $y$ with period $2K(k_2)/m_2$. In general,
$\theta_1(x+2K(k_1)/m_1)\neq \theta_1(x)+2 n \pi$, for some integer $n$. Thus,
the solution \rf{eqn:ans} is  usually not periodic in $x$ or in $y$. Imposing
this periodicity requires a quantization of the phase $\theta_1(x)$ and
$\theta_2(y)$~\cite{becpre1}. 
It is unclear if such a quantization is possible, since only one
free parameter is available to satisfy two conditions. There are two special
cases in which phase quantization is not a concern. The first case results in
trivial-phase solutions, for which $c_1=0$, $c_2=0$. The second case is the
trigonometric limit, in which $k_1=0$, $k_2=0$. 

The solution has trivial phase in the $i$-direction if $c_i$ is zero, where
$i=1~(2)$ corresponds to the $x~(y)$-direction. There are three possibilities: 
\alpheqn
\bea
B_i&=&0: ~~~\,~~~~~~~ r_i=\sqrt{A_i}~\sn(m_i x,k_i),\\
B_i&=&-A_i: ~~\,~~~~ r_i=\sqrt{-A_i}~\cn(m_i x,k_i),\\\la{eqn:dn}
B_i&=&-A_i/k_i^2: ~~ r_i=\sqrt{-A_i/k_i^2}~\dn(m_i x,k_i),
\eea
\resetalpheqn
where $\cn(m_i x_i,k_i)$ is the Jacobian elliptic cosine function, and $\dn(m_i
x_i, k_i)$ denotes the third Jacobian elliptic function. 

The solution is trigonometric in the $i$-direction if $k_i$ is zero. Then
\beq\la{eqn:trig}
r_i^2(x)=A_i\sin^2(x_i)+B_i, ~~
\tan(\theta_i)=\sqrt{1+\frac{A_i}{B_i}}\tan(m_i x_i),
\eeq
and phase quantization is satisfied. 
Notice that it is possible for the solution to have trivial phase in one
direction and be trigonometric in the other. 

The main stability results from \cite{becprl1,becpre1} can be generalized to
two dimensions in a straightforward manner. The details of this will be
presented elsewhere. Here a brief summary is given. Examining the linear
stability of exact trivial-phase stationary solutions $\psi(x,y,t)=
r(x,y)\exp(-i \omega t)$ of equation
\rf{eqn:nls2d} gives rise to the eigenvalue problem
\beq
\left(
\ba{cc}
0&-L_-\\L_+&0
\ea
\right)\left(
\ba{c}
u_1\\u_2
\ea
\right)=\lambda\left(
\ba{c}
u_1\\u_2,
\ea
\right)
\eeq
where the linear Schr\"odinger operators $L_+$ and $L_-$ are
\alpheqn
\bea
L_+&=&-\frac{1}{2}\left(\p_x^2+\p_y^2\right)+3r^2(x,y)+V(x,y)-\omega,\\
L_-&=&-\frac{1}{2}\left(\p_x^2+\p_y^2\right)+r^2(x,y)+V(x,y)-\omega.
\eea
\resetalpheqn
Reasoning similar to the arguments in \cite{becprl1,becpre1} leads to the
following conclusions:  
\begin{itemize}

\item If $r(x,y)>0$, then $r(x,y)$ is the ground state of $L_-$ and
$\psi(x,y,t)$ is linearly stable. 

\item If $r(x,y)$ has any zeros, it is not the ground state of $L_-$. If in
addition $L_+$ is a positive operator, then $\psi(x,y,t)$ is unstable. 

\item In all other cases, no conclusion is obtained from this stability
analysis. 

\end{itemize}

Primarily, this analysis immediately establishes the linear stability of
a solution $\psi(x,y,t)=r_1(x) r_2(y) \exp(-i \omega t)$, where both $r_1(x)$
and $r_2(y)$ are strictly positive. Thus, the solution with both the $x$- and
$y$-dependence specified by \rf{eqn:dn} is linearly stable.

\section{Numerical Results}\la{sec:num}

In this section, the result of numerically solving \rf{eqn:nls2d} with initial
conditions chosen from the exact solutions given in the previous section are
discussed.  The numerical procedure uses a fourth-order Runge-Kutta method in
time and a filtered pseudospectral method in space.  For each experiment, a
small amount of white noise was added to the initial conditions as a
perturbation.

Consider the solution
\beq\la{eqn:dnsol}
\psi(x,y,t)=\frac{\sqrt{-A_1}\sqrt{-A_2}}{k_1 k_2}\dn(m_1 x,k_1)
\dn(m_2 y,k_2)e^{-i\omega t}.
\eeq
where $A_1<0$ and $A_2<0$.
%
%
\begin{figure}[tb]
\centerline{\psfig{figure=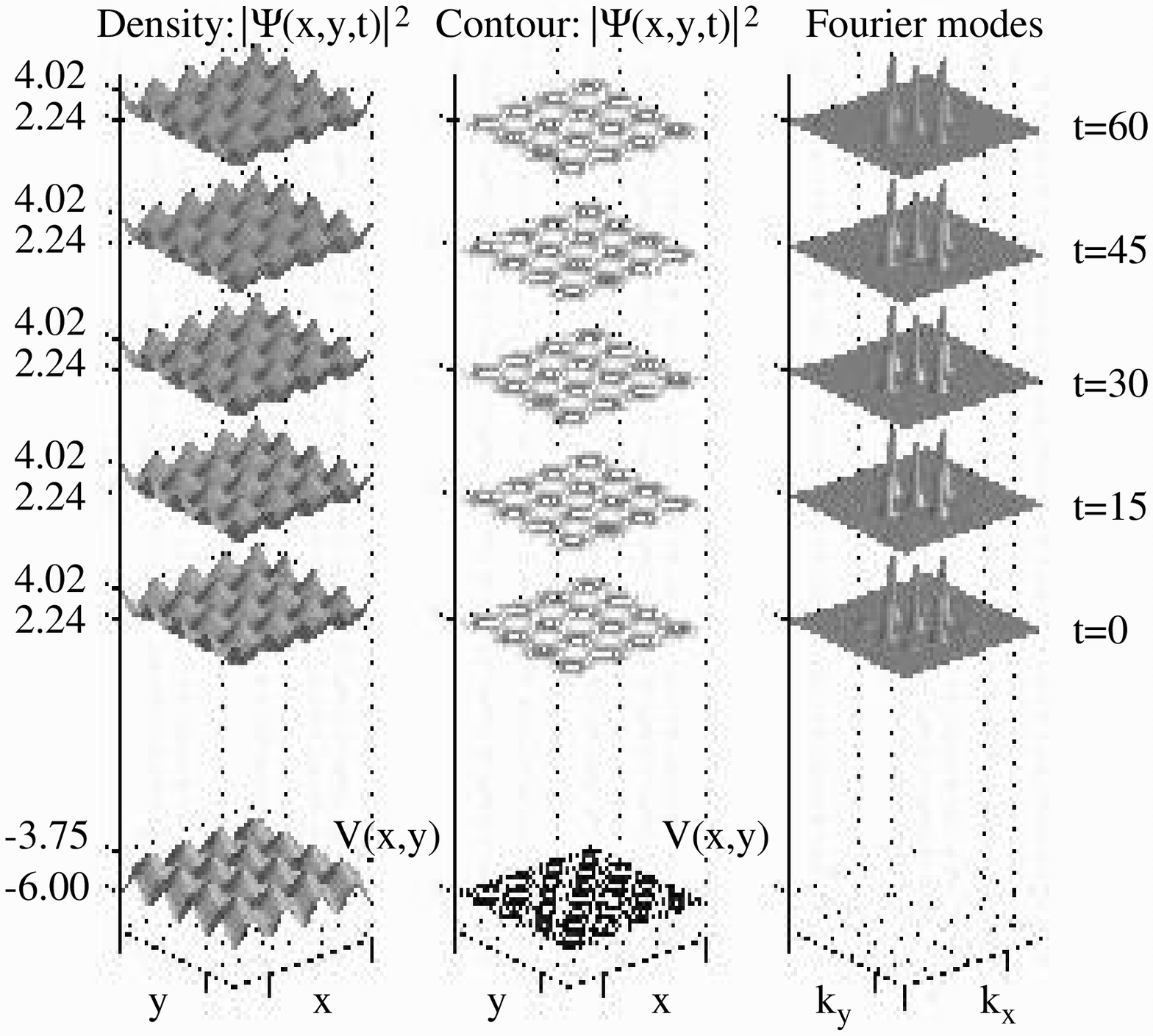,width=140mm,silent=}}
\begin{center}
\caption{\la{fig:dn} The stable solution from Eqn. \rf{eqn:dnsol}. Here
$k_1=k_2=1/2$, $m_1=m_2=1$ and $A_1=A_2=-1$.}
\end{center}
\end{figure}
%
%
As stated in the previous section, this solution is linearly stable. This is
confirmed by the numerics presented in Fig. \ref{fig:dn}. The three different
columns give from left to right the dynamics of the solution, the same using
contour plots, and the arctan  of the Fourier power spectrum of the solution. In
the first two columns, the bottom figure shows the potential $V(x,y)$. 
The stability of this solution is reminiscent of the stability of a plane
wave solution of the two-dimensional defocusing nonlinear Schr\"odinger
equation~\cite{sulem}.

Next, consider the trivial-phase solution
\beq\la{eqn:cnsol}
\psi(x,y,t)=\sqrt{-A_1}\sqrt{-A_2}~\cn(m_1 x,k_1)\cn(m_2 y,k_2)e^{-i
\omega t}.
\eeq
%
%
\begin{figure}[tb]
\centerline{\psfig{figure=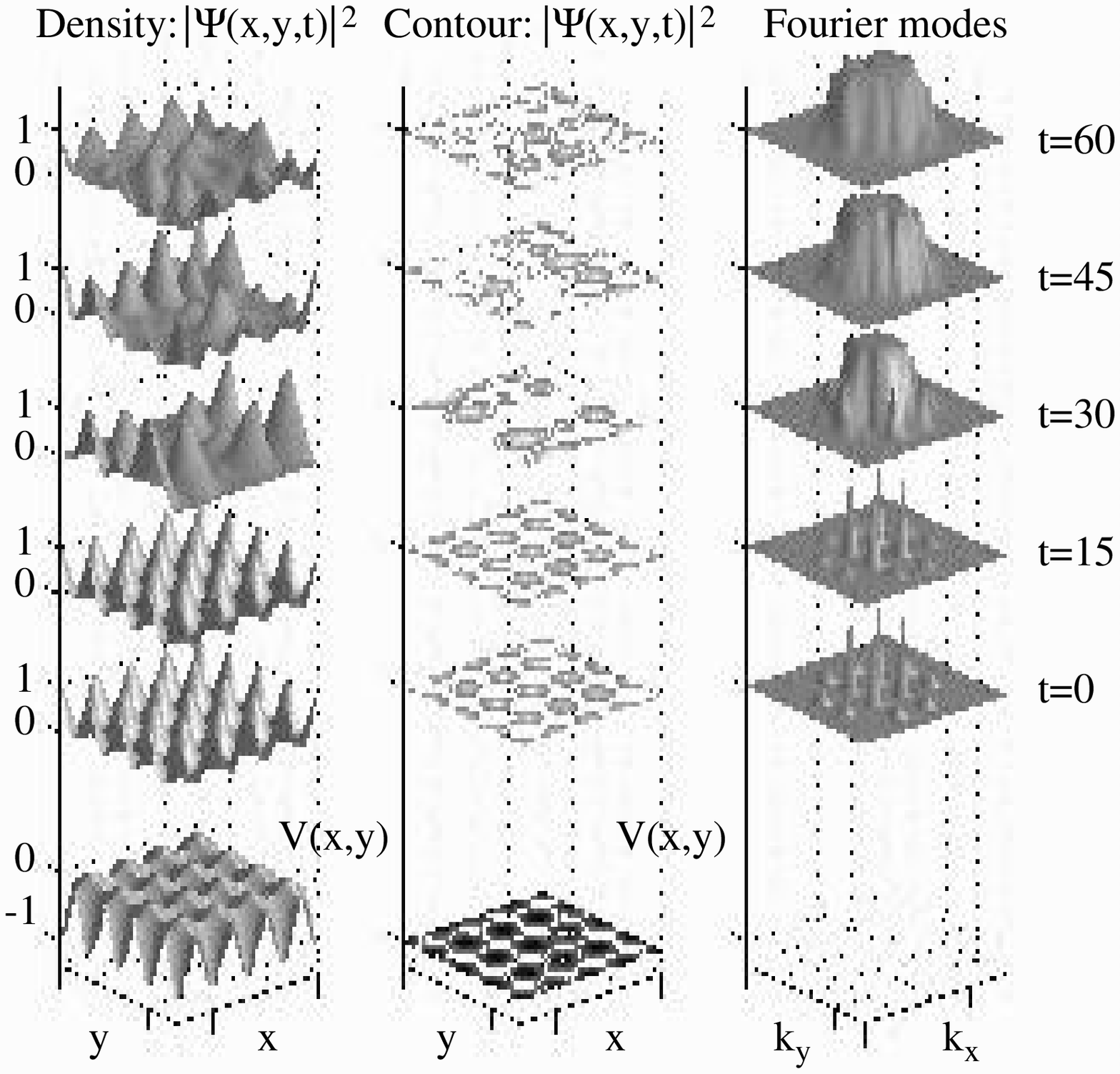,width=140mm,silent=}}
\begin{center}
\caption{\la{fig:cn} The unstable solution from Eqn. \rf{eqn:cnsol}. Here
$k_1=k_2=1/2$, $m_1=m_2=1$ and $A_1=A_2=1$. }
\end{center}
\end{figure}
%
%
The numerics of Fig. \ref{fig:cn} shows that this solution is unstable. For the
equation \rf{eqn:nls2d}, the $L^2$-norm of the solution is conserved, hence so
is the $L^2$-norm of its Fourier spectrum. This is not reflected in Fig.
\ref{fig:cn}, due to the arctan transformation, which is used to
diminish the range of the power spectrum. The onset of instability occurs
between $t=15$ and $t=30$. A detailed look at this onset is given in Fig.
\ref{fig:explosion}. 
%
%
\begin{figure}[tb]
\centerline{\psfig{figure=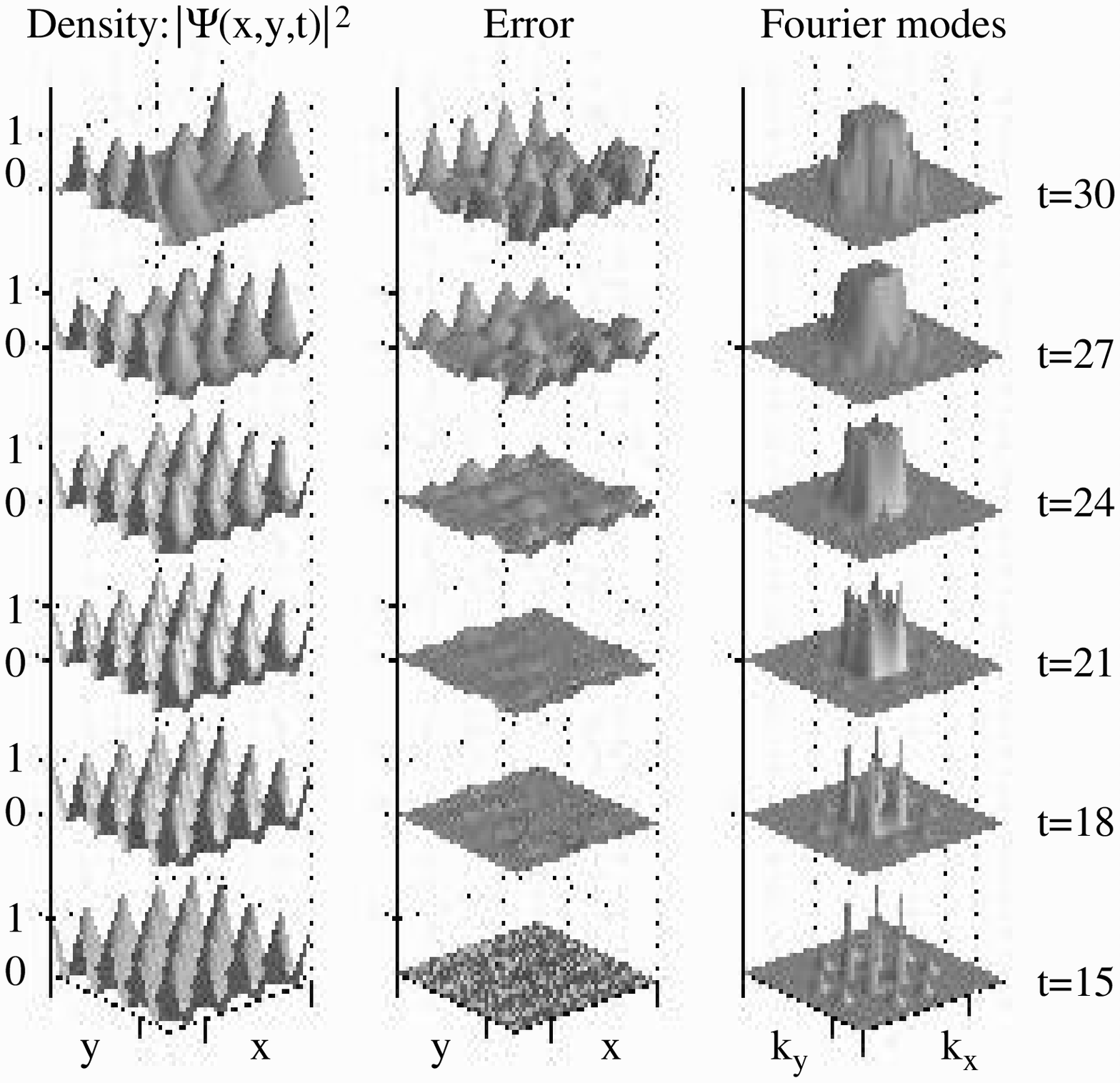,width=140mm,silent=}}
\begin{center}
\caption{\la{fig:explosion} A blow-up of the onset of instability occurring in
Fig. \ref{fig:cn}. The middle figure shows $||\psi(x,y,t)|^2-
|\psi(x,y,t=0)|^2|$}.
\end{center}
\end{figure}
%
%
The middle column of this figure shows $||\psi(x,y,t)|^2-|\psi(x,y,t=0)|^2|$.
The instability begins to develop around $t=18$. Before the instability occurs,
$|\psi(x,y,t)|^2$ is equal to its initial condition, up to effects due to the
added noise and numerical round-off, as is expected, since $\psi(x,y,t)$ is
an exact stationary solution of \rf{eqn:nls2d}. The solution 
\beq\la{eqn:snsol}
\psi(x,y,t)=\sqrt{A_1 A_2}\sn(m_1 x,k_1)\sn(m_2 y,k_2)e^{-i \omega t}.
\eeq
is observed numerically to be unstable as well. Its dynamics is very similar to
that of the solution \rf{eqn:cnsol}.  These results are consistent with
the one-dimensional results obtained previously~\cite{becprl1,becpre1}.

Analytical results for the stability of the 
nontrivial phase solutions are difficult to obtain, and so
far numerical investigations are the only means of examining these solutions.
Since phase quantization is a serious complication for the numerics, it is
convenient to work with solutions \rf{eqn:trig} that are trigonometric in both
directions since phase quantization is automatically satisfied. 
Thus, we consider solutions of the form 
%
%
\beq\la{eqn:trigsol}
\!\!\!\!\!\!\!\!\!\!\!
\psi(x,y,t)=\sqrt{\left(A_1\sin^{2}(m_1 x)+B_1\right)\left(
A_2\sin^{2}(m_2 y)+B_2\right)}e^{i\theta_1(x)+i 
\theta_2(y)-i \omega t},
\eeq
with
\beq \la{eqn:trigphase}
\tan(\theta_1(x))=\sqrt{1+\frac{A_1}{B_1}}\tan(m_1 x), ~~
\tan(\theta_2(y))=\sqrt{1+\frac{A_2}{B_2}}\tan(m_2 y). 
\eeq
These solutions are stable
or unstable depending on the offset parameters $B_1$ and $B_2$.
As shown in Fig.~\ref{fig:dn} the off-set $\dn(m_ix_i, k_i)$-type solution is
stable whereas in Fig.~\ref{fig:cn} the unstable $\sn(m_i x_i, k_i)$ and $\cn(m_i x_i,
k_i)$ solutions without offset are unstable.  These trivial phase solutions
suggest that offset is essential for stability.   In Fig.~\ref{fig:ntp-unstable},
$A_1=A_2=1$ and $B_1=B_2=0.5$ and the onset of
instability of the nontrivial phase solution is observed.  
Here the chosen offset ($B_1=B_2=0.5$) 
is insufficient to stabilize the dynamics.  In contrast,
Fig.~\ref{fig:ntp-stable} has parameter values $A_1=A_2=1$ and $B_1=B_2=1$ and
is stable.  Thus with a higher offset, the nontrivial phase solution
is stabilized.  The boundary between the stable and unstable regions is
difficult to calculate analytically and we rely on numerical simulations to
determine the amount of offset required for stability.  This behavior is
again consistent with the one-dimensional case~\cite{becpre1}.

%
\begin{figure}[tb]
\centerline{\psfig{figure=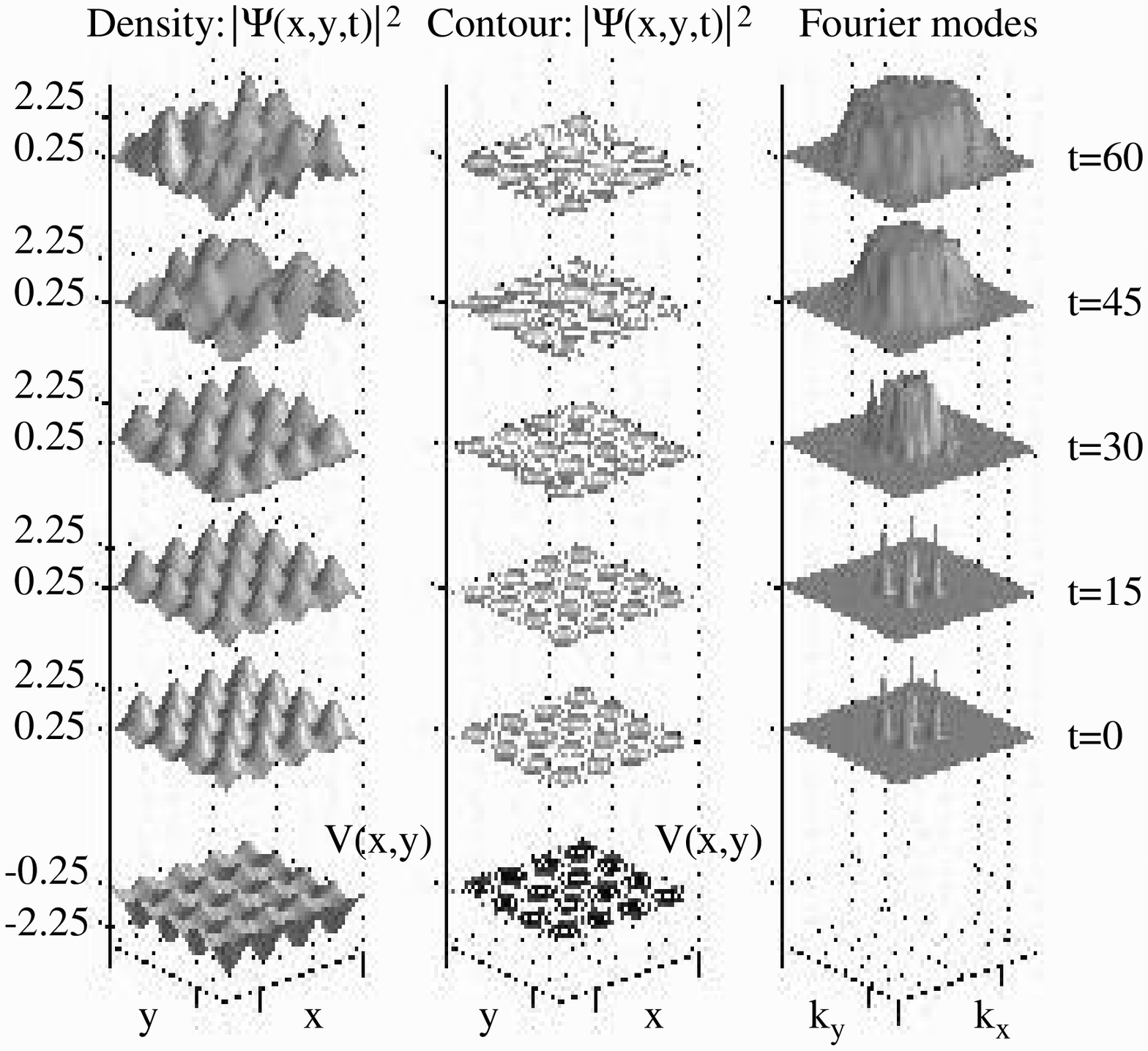,width=140mm,silent=}}
\begin{center}
\caption{\la{fig:ntp-unstable}  An unstable solution from Eqns. \rf{eqn:trigsol}
and \rf{eqn:trigphase}. Here
$k_1=k_2=1/2$, $m_1=m_2=1$, $A_1=A_2=1$, and $B_1=B_2=0.5$.}
\end{center}
\end{figure}
%

%
\begin{figure}[tb]
\centerline{\psfig{figure=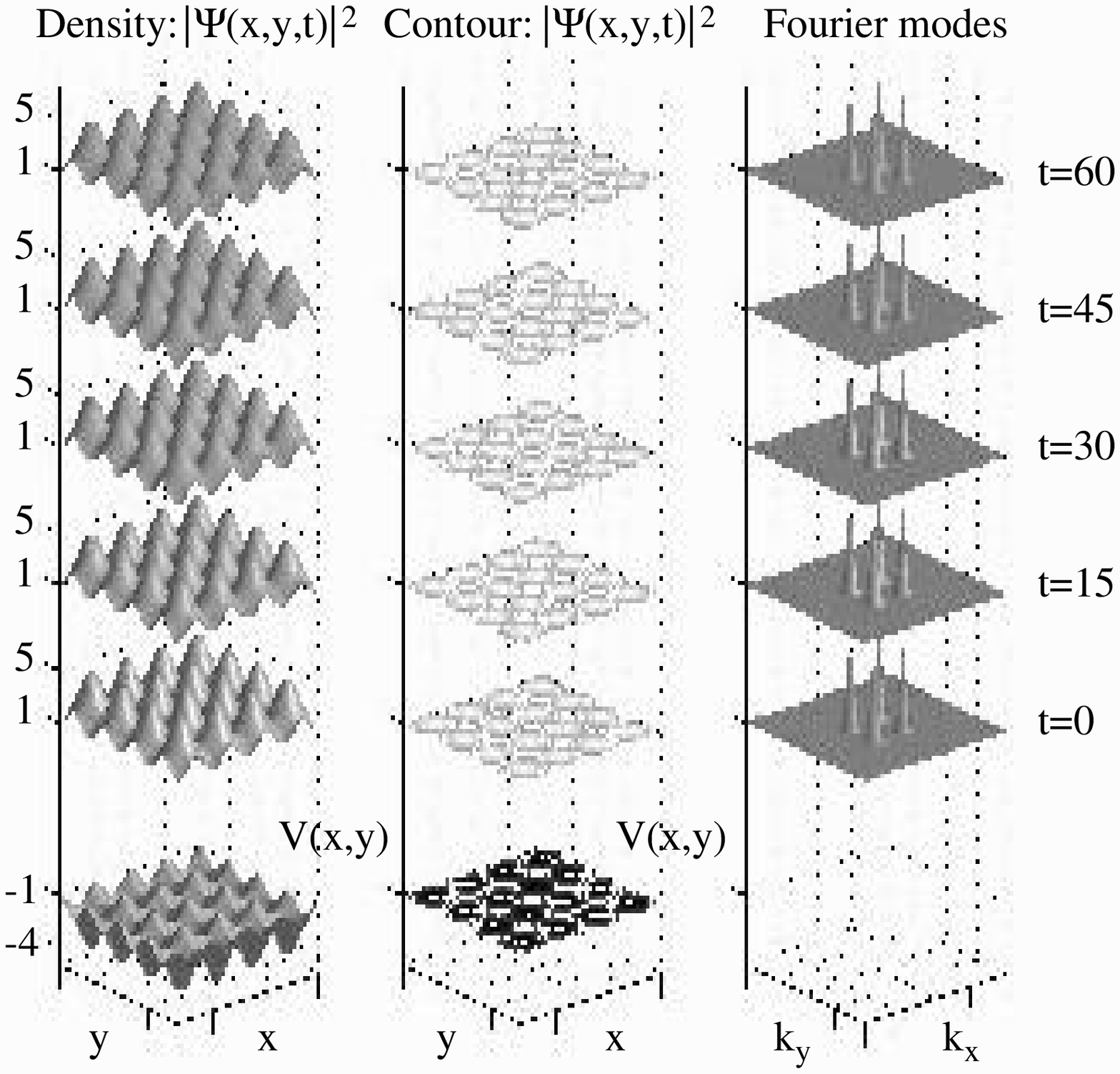,width=140mm,silent=}}
\begin{center}
\caption{\la{fig:ntp-stable} A stable solution from Eqns. \rf{eqn:trigsol}
and \rf{eqn:trigphase}. Here
$k_1=k_2=1/2$, $m_1=m_2=1$, $A_1=A_2=1$, and $B_1=B_2=1$.}
\end{center}
\end{figure}
%

\section{Summary and Conclusions}

We considered the repulsive nonlinear Schr\"odinger equation with an elliptic
function potential in two dimensions.
Periodic solutions of this
equation were found and their stability was investigated both analytically and
numerically.  Using analytical results for trivial-phase solutions, we showed
that solutions with sufficient offset are linearly stable.  
This is confirmed with extensive numerical
simulations on the governing nonlinear equation.  Likewise, nontrivial-phase
solutions are stable if they are sufficiently off-set.

Since our equations is a model for a Bose-Einstein condensate trapped in a lattice
potential, our results imply that a large number of condensed atoms is sufficient
to form a stable, periodic condensate.  Physically, this implies stability of
states near the Thomas--Fermi limit.  The results are consistent with the
one-dimensional trapping of a BEC condensate in a standing light wave.
To quantify this phenomena, we consider the $k=0$ limit and note that in the
trigonometric limit $k\rightarrow 0$ the number of particles $n$ per potential
well is given by 
$n=\left< |\psi(x,y,t)|^2 \right> = \left< r_1^2(x)\right>\left< r_2^2(y) 
\right> = (A_1/2+ B_1)(A_2/2+B_2)$, where $\left< \cdot \right>$ denotes
the spatial average.  In the context of
the BEC, and for a fixed atomic coupling strength, this means a large number of
condensed atoms $n$ per potential well is sufficient to provide an offset on the
order of  the potential strength.  This ensures stabilization of the
condensate. Alternatively, a condensate with a large enough number of atoms can
be  interpreted as a developed condensate for which the nonlinearity  acts as a
stabilizing mechanism. 
   
{\bf Acknowledgments:} The authors acknowledge useful conversations with J. C.
Bronski, L. D. Carr, R. Carretero-Gonz\'alez, K. Promislow, and W.
Reinhardt.

\bibliographystyle{unsrt}

\bibliography{bec2D}

\end{document}